\documentclass[
twocolumn,aps,prr,superscriptaddress
]{revtex4-2}

\usepackage{graphicx,color,hyperref}

\usepackage{amsmath}
\usepackage{amsfonts}
\usepackage{amssymb}
\usepackage{dsfont}

\usepackage{braket}

\newcommand{\tr}{\mathrm{tr}}

\newcommand{\pd}{{\vphantom{\dagger}}}
\usepackage[normalem]{ulem} 

\begin{document}

\title{
Dynamics of measured many-body quantum chaotic  systems
}

\author{Alexander Altland}
\affiliation{Institut f\"ur Theoretische Physik, Universit\"at zu K\"oln, Z\"ulpicher Str. 77, 50937 Cologne, Germany
}

\author{Michael Buchhold}
\affiliation{Institut f\"ur Theoretische Physik, Universit\"at zu K\"oln, Z\"ulpicher Str. 77, 50937 Cologne, Germany
}

\author{Sebastian Diehl}
\affiliation{Institut f\"ur Theoretische Physik, Universit\"at zu K\"oln, Z\"ulpicher Str. 77, 50937 Cologne, Germany
}

\author{ 
Tobias Micklitz
}
\affiliation{Centro Brasileiro de Pesquisas F\'isicas, Rua Xavier Sigaud 150, 22290-180, Rio de Janeiro, Brazil 
} 

\date{\today}


\begin{abstract}
We consider the evolution of continuously measured many-body chaotic  quantum
systems. Focusing on the dynamics of state purification, we analytically describe the
limits of strong and weak measurement rate, where the latter case is challenging in that  monitoring up to
time scales exponentially long in the numbers of particles is required. We complement
the analysis of the limiting regimes with the construction of an effective replica
theory providing information on the stability and the symmetries of the respective
phases. The analytical results are tested by comparison to exact diagonalization.
\end{abstract}
\maketitle

\emph{Introduction ---} Continuous time or repeated projective measurements performed on complex quantum
systems may trigger a \textit{measurement-induced quantum phase
transition}~\cite{Skinner2019,Fisher2018,Li2019b,gullans2019,choi2020prl,fan2020selforganized,nahum2021prxq, Barkeshli2021,doggen2021generalized, bao2021symmetry,jian2021quantum,turkeshi2021measurementinduced,ippoliti2020,Zabalo2020,zabalo2021operator,lifisher2021,Jian2020,jian2021syk, buchhold2021effective,alberton2021enttrans,minato2021fate,block2021measurementinduced,muller2021measurementinduced,Biella202,Romito2020,Schomerus2019,fuji2020,boorman2021diagnosing}. What sets this transition apart from generic phase transitions is that it remains invisible in system density operators averaged over measurement-detector degrees of freedom. It is, rather, of statistical nature and manifests itself through correlations of individual ``quantum trajectories'' traced out by  a system
subject to repeated monitoring with random outcomes. Observables serving
as effective order parameters include R\'enyi or von Neumann entanglement entropies~\cite{Skinner2019,Fisher2018,lunt2020,Shengqi2021,Vedika2021,Nahum20a}, or the purity of the evolving quantum states~\cite{Gullans2020, Monroe2021,gullans2019}. They all have in common that they are
expressed through moments or \emph{replicas} of the system's density operator~\cite{alberton2021enttrans,buchhold2021effective,Hsieh2021,Hsieh2021a}.

The necessity to deal with  system replicas complicates the theoretical description
of measurement
dynamics~\cite{chen2020,Jian2020,li2020conformal,zhang2021syk,jian2021syk,jian2020criticality,li2021statistical,lifisher2021,Bao_2020,buchhold2021effective}.
However,  external monitoring  also implies a simplification: A continuously observed
system is subject to \emph{noise} representing the randomness of  measurement
outcomes~\cite{Cao2019,minoguchi2021continuous,alberton2021enttrans,Carlton1987,Kyrylo2021}.
Decoherence due to this noise effectively projects states onto configurations
diagonal in the measurement basis. For nonintegrable systems   the repeated
projection actually simplifies the dynamics  compared to that of the unmeasured
system, and it is this principle that allows us to gain traction with the problem.

In contrast to unitary quantum circuits, the measurement-induced dynamics of
\emph{nonintegrable Hamiltonian} systems is still largely unexplored with only few
available numerical results~\cite{lunt2020,doggen2021generalized,fuji2020}. In this
paper, we focus on this system class, for particle numbers, $N$, large but finite, as
relevant to quantum hardware in current technological
reach~\cite{Gullans2020,Bentsen2021,Gopa2021}. Conceptually, our main goal is the
construction of analytical approaches versatile enough to describe the dynamics of
such systems in different regimes. Specifically, we will find
that the cases of weak and strong measurement  call for individual treatments,
tailored to the dominance of ergodic chaotic time evolution and repeated measurement
intrusion, respectively. These limiting cases are separated by a symmetry breaking
phase transition whose presence and parametric dependence on system parameters we
describe in terms of a semi-phenomenological replica mean field theory. Exact
diagonalization shows that results obtained in this framework  enjoy a high level of
stability away from the limits in which they were obtained. In this way, the present
three-thronged approach  describes the different manifestations of monitored
evolution in quantum ergodic systems of mesoscopic extension under reasonably general
conditions.

\noindent \emph{Model ---} We consider a system with $N\gg 1$ fermion states
$\alpha=1,\ldots N$ governed by the Hamiltonian, $\hat
H=\sum_{\alpha,\beta}c_\alpha^\dagger h_{\alpha\beta} c_\beta + \hat H_\mathrm{int}$,
where $\hat H_\mathrm{int}$ is a two-body interaction. Concerning the free part,
$h_{\alpha\beta}$, we need not be specific, other than that it is chaotic with
extended single particle states $|\psi_i\rangle $. An expansion of the Hamiltonian in
the single particle eigenbasis brings it into the form $\hat H=\sum_\alpha
c_i^\dagger c_i\epsilon_i + \sum_{ijkl} J_{ijkl}c_i^\dagger c_j^\dagger c_k c_l$.
Reflecting the effective randomness of chaotic wave functions, the interaction matrix
elements may be considered as stochastic variables~\cite{Altshuler2000} with variance $\langle
|J_{abcd}|^2\rangle_J \equiv 6 J^2/(2N)^3$. Depending on the relative strength of the
interaction and the single particle contributions this model may be in one of two phases~\cite{Altland2019}: for
single particle band widths $W>J$ it defines a Fermi liquid with quasi particle
states renormalized by interactions. In the opposite case, strong interactions send
it into a non-Fermi liquid phase with the characteristics of a `strange metal'.  As
we will see, the results of our analysis are largely insensitive to this distinction,
and therefore enjoy a considerable level of universality.

 To 
 simplify the model somewhat, we 
sacrifice particle number conservation: introducing real (Majorana) fermions through
$c_i = \frac{1}{2}(\chi_{2i-1}+ i \chi_{2i})$, we generalize the interaction to
$\hat{H}_\mathrm{int}=\sum_{a,b,c,d=1}^{2N}J_{abcd}\chi_a \chi_b \chi_c \chi_d$, where the real  constants $J_{abcd}$ are implicitly defined by the complex $J_{ijkl}$. This generalization puts us into the class of the Majorana SYK$_{2+4}$-model containing maximally random two and four fermion operators. Compared to the complex version,   the many body chaotic dynamics now  mixes between all states in the $2^N$-dimensional fermion Fock space (and not just sectors of definite particle number).


Our observable of interest will be
the purity $\langle \tau_t\rangle \equiv \langle \mathrm{tr}(\rho_t^2)\rangle$, where
$\langle...\rangle$  denotes the average over measurement
runs~\cite{gullans2019,Gullans2020,Monroe2021}. This quantity
indicates the transition between phases with weak and strong measurement rate through
its time dependence: the typical time scale $ t_\mathrm{p}$ to reach asymptotic
purification, $\tau_t \to 1$ for weak (strong) measurement is exponentially long $
t_\mathrm{p}\sim \exp(N)$ (logarithmically short $ t_\mathrm{p}\sim \log(N)$) in the
system size $N$. We will discuss how these limits are realized, and discuss the
stability range of the respective time dependencies. For simplicity, we consider the pure interaction model, $W=0$, in the main text. The numerical analysis of the generalized model in the supplemental material leads to no significant changes in the results. 

\begin{figure}[t!]
\centering
\includegraphics[width=8.5cm]{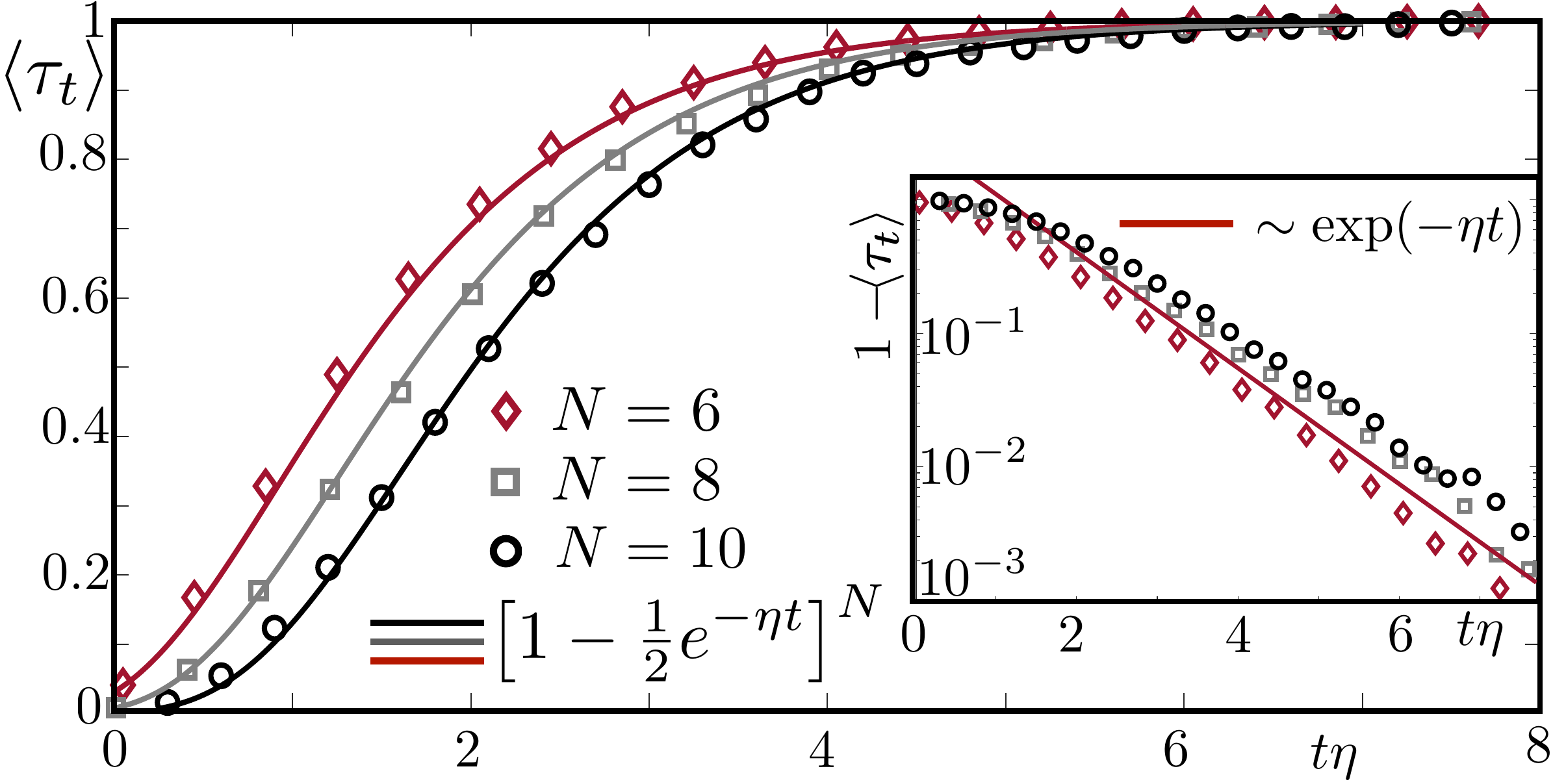}
\caption{\label{fig:PurSmallJ}
Purification ($\langle\tau_t\rangle=\langle\text{tr}\rho_t^2\rangle$) of a fully mixed initial state $\rho_0\sim\mathds{1}$ for strong measurements, $J=10^{-2} \eta$, and different system sizes $N$. Discrete dots are averages over $500$ simulated trajectories. Bold lines correspond to the analytical strong measurement prediction $\langle \tau_t\rangle=(1-\frac{1}{2}e^{-\eta t} )^N$.}
\end{figure}

\noindent \emph{Strong measurement ---} We consider measurement so strong that the
scrambling effect of $\hat H$ on states projected onto the occupation number
eigenbasis $|n\rangle = \otimes^N |n_i\rangle $ is negligible. In this limit,
individual of the qubit states defined by $n_i=0,1$ can be considered separately.
Assuming an initially fully mixed state, $\rho_0=\sum_n |n\rangle \langle n|$, the
density matrix remains diagonal in the occupation number basis, and $\tau_t=
\tr(\rho_t^2)= \tau_{1t}^N$ factorizes into the $N$th power of single qubit purities,
$\tau_{1t}$. To describe the evolution of the latter, we assume that each qubit is
measured with an average rate $\eta$. The probability $p$ that no measurement has taken place
after time $t$ then is $e^{-t\eta}$. In this case, the qubit remains fully mixed and $ \tau_1=1/2$, otherwise the qubit
state is known and $\tau_1=1$. We thus obtain $\langle\tau_{1t}\rangle= p\frac{1}{2}+(1-p)=1-\frac{1}{2}e^{-\eta t}$, and $\langle \tau\rangle=(1-\frac{1}{2}e^{-\eta t} )^N$. 
For times exceeding the measurement time,
$t>\eta^{-1}$, we may approximate $\langle \tau_t\rangle\approx
\exp(-\frac{N}{2}e^{-\eta t})$, showing that $ t_{\mathrm{p}}\equiv \eta^{-1}\ln N$
sets the characteristic time scale at which purification is reached. Finally, a simple replacement $1/2\rightarrow (1/2)^r$ yields the $r$th moments of the purity, and from there the \emph{typical} purity $\tau_{\mathrm{typ},t}\equiv \exp\langle  \ln \tau_t\rangle=\exp \left[\partial_r\langle  \tau_t^r\rangle\right]_{r=0}$ as $\tau_{\mathrm{typ},t}=\exp(- N e^{-\eta t}\ln 2)$, showing that the strong measurement purity essentially is a self averaging quantity. 
Figure~\ref{fig:PurSmallJ} shows that these predictions match the results of exact numerical simulations performed
for a continuous time measurement protocol (see~\cite{supplement}) at $J/\eta = 10^{-2}$. 

\noindent \emph{Weak measurement ---} The analysis of the weak measurement regime is more challenging. We consider measurement rates $\eta\ll J$ much smaller than the inverse of the time scale $\sim J^{-1}$ at which the SYK dynamics approaches ergodicity. In this case we anticipate that the information $\ln 2$ learned by measuring a single qubit is 
scrambled over the entire Hilbert space 
between two measurement events. The goal is to
describe how a tiny fraction of this information is retained and a purified state reached,
albeit on very large time scales. Referring to the supplemental material for more details, we represent the density operator
after a sequence of $l$ projective qubit measurements as $\rho_l = \mathcal{N}_l Z_l$
with $\mathcal{N}_l=\mathrm{tr}(Z_l)^{-1}$ and $Z_l=P_l U_l P_{l-1} \ldots
P_{l-1}U_l^\dagger P_l=P_l U_l Z_{l-1}U_l^\dagger P_l$ in a recursive definition.
Here, $P_k$ are projectors onto a definite state $0,1$ of any of the $N$-qubits
(which one does not matter) and $U_k$ are $D\times D$ dimensional unitary matrices,
assumed independently Haar distributed. These operators serve as proxies to the
ergodic dynamics, and their independent distribution reflects the randomly
distributed times between measurements.
 The purity
after $l$ measurements is given by $\langle
\mathcal{N}_l^2\mathrm{tr}(Z_l^2)\rangle$. We evaluate this expression under the
additional assumption of approximate statistical independence of the normalization factor and
the operator trace $\langle \tau_l\rangle\approx \langle
\mathcal{N}_l^{-2}\rangle^{-1}\langle \mathrm{tr}(Z_l^2)\rangle$. This approximation is not backed by a small parameter and its legitimacy must be checked by comparison to exact diagonalization. 

Defining $X_{2l}\equiv \langle\tr(Z_l^2)\rangle$, and $X_{1l}\equiv \langle(\tr(Z_l))^2\rangle$ the recursive computation of the purity is now reduced to that of the matrix averages $X_{2l} = \langle  \mathrm{tr}([P_lU_l Z_{l-1} U^\dagger_l P_l]^2 )\rangle$ and $X_{1l}=\langle (\mathrm{tr}(P_l U_l
Z_{l-1} U^\dagger_l P_l ))^2\rangle$. The Haar averaged products of four matrices can be computed in closed form (see~\cite{supplement}) with the simple result $X_l=\left(\begin{smallmatrix}
         z_1&z_2\cr
         z_2&z_1
     \end{smallmatrix} \right)X_{l-1}$,
where $X_l=(X_{1l},X_{2l})^T$, and $z_1\approx 1/4$ and $z_2\approx 1/D$. This
equation describes the evolution of the purity in terms of just two trace invariants
$X_{1l,2l}$. Its structure reflects the general principle mentioned in the
introduction: Chaotic mixing implies that only universal trace
invariants survive at time scales exceeding the ergodicity
time.

\begin{figure}[t!]
\centering
\includegraphics[width=7.5cm]{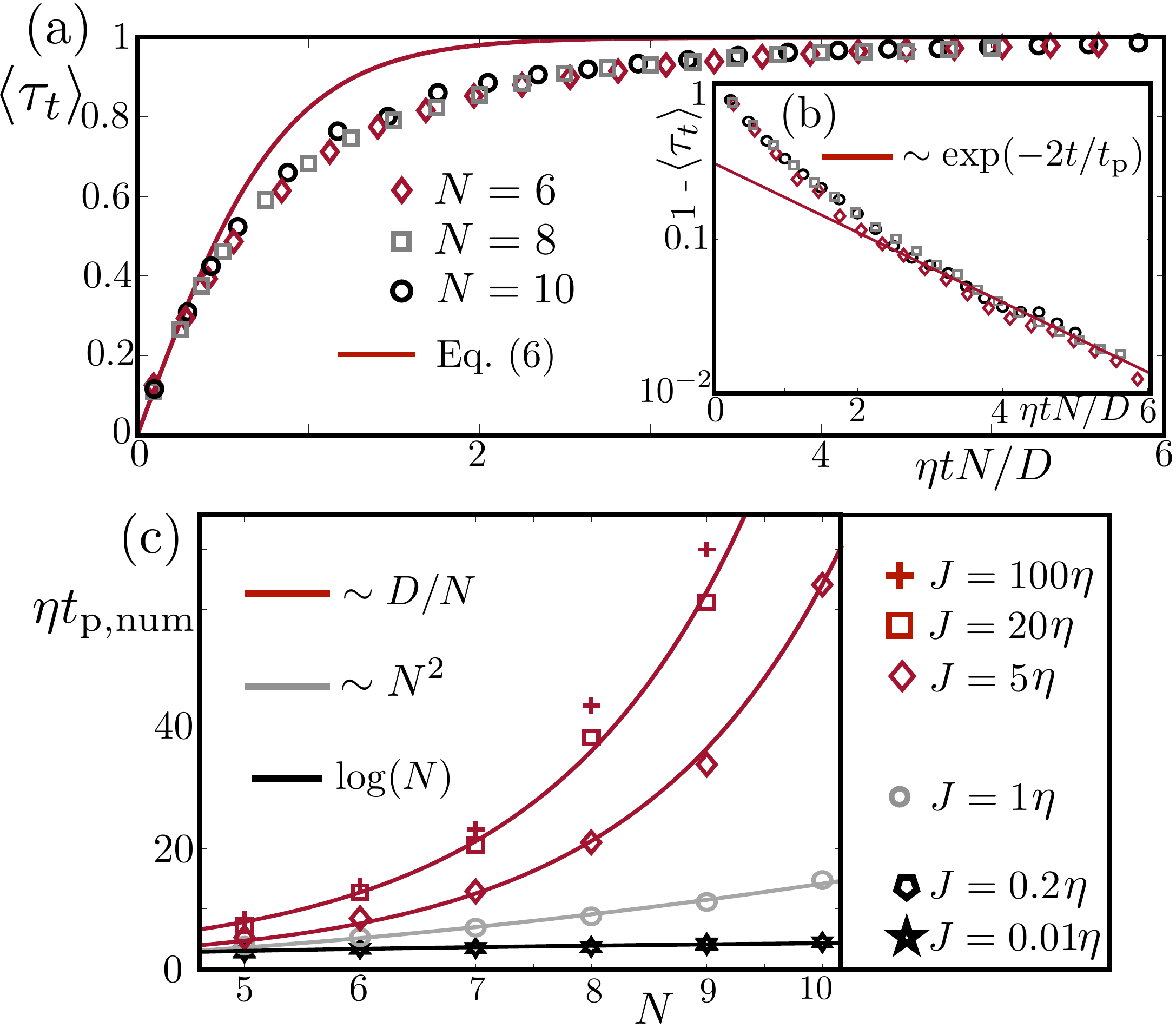}
\caption{\label{fig:PurLargeJ}
(a)+(b) Purification of the weakly measured system ($J=5\times 10^3 \eta$) compared with the analytical prediction Eq.\eqref{eq:LargeJPurTime}. The dots are obtained by averaging over $500$ numerically simulated trajectories. A scaling collapse of $\tau_t=\text{tr}\rho^2_t$ is obtained when evolving it in the dimensionless time $\eta t N/D$. (c) Purification time $ t_{\mathrm{p},\mathrm{num}}$ (defined by $\tau_{ t_{\mathrm{p},\mathrm{num}}}=0.9$) for different values of $J/\eta$. The purification time matches the weak measurement prediction (Eq.~\eqref{eq:LargeJPurTime}, red data) for $J\ge5\eta$ and the strong measurement prediction ($ t_{\mathrm{p},\mathrm{num}}\eta\sim\log(N)$, black data) for $J\le0.2\eta$. The scaling $ t_{\mathrm{p},\mathrm{num}}\sim N^2$ around $J=\eta$ is associated with the vicinity of the entanglement transition, expected at $J\simeq 2\eta$ according to our theory.}
\end{figure}

The linear recursion relation is straightforwardly solved by an exponential ansatz subject to the initial condition $X_{20}=\tr(\rho_0^2)=D^{-1}$ and $X_{10}=(\tr(\rho_0))^2=1$. Noting that the step number $l=t/\eta N$ equals physical time divided by the the total measurement rate, we obtain the  purity   $\langle \tau_l\rangle = X_{2l}/X_{1l} $ as 
\begin{align}\label{eq:LargeJPurTime}
    \langle \tau_t\rangle\approx \frac{\sinh\left(\frac{t}{ t_\mathrm{p}}\right)+D^{-1}\cosh\left(\frac{t}{ t_\mathrm{p}}\right)}{\cosh\left(\frac{t}{ t_\mathrm{p}}\right)+D^{-1}\sinh\left(\frac{t}{ t_\mathrm{p}}\right)}.
\end{align}
This result predicts purification, $\langle \mathrm{tr}(\rho^2_{\infty})\rangle =1$,
at  $t\sim t_\mathrm{p}=D/N\eta$ no matter how  small the measurement rates.
However, the  purification time scale, $ t_\mathrm{p}$, now grows exponentially in  the number of qubits, in contrast to the logarithmic scaling $
t_\mathrm{p}\sim\ln (N)/\eta$ in the strong measurement regime.
Figure~\ref{fig:PurLargeJ} compares this prediction to numerics  for
$J/\eta=5\times 10^3$ and different system sizes, $N=6,8,10$. We indeed find data collapse for the  scaled variable $t/
t_\mathrm{p}$. For intermediate times ($\eta t N\approx D)$,
Eq.~\eqref{eq:LargeJPurTime} overestimates the purification with a maximum error of
$10\%$ --- likely a consequence of a partial violation of the above assumptions on statistical independence.

The bottom panel of Fig.~\ref{fig:PurLargeJ} compares the  purification time
$t_{\mathrm{p},\mathrm{num}}$, here defined as the time scale at which the purity has
reached the value $0.9$, to the analytical predictions. It turns out that  the two
time dependencies $\eta t_\mathrm{p}\sim  \ln N$ and $\eta t_\mathrm{p}\sim D/N$ for
$\eta/J\gg 1 $ and $\eta/J\ll 1$, respectively, show a remarkable degree of stability
away from the limits in which they were obtained. Hinting at the existence of a phase
transition,  they cover almost the entire parameter axis $\eta/J$, except for a range
$0.2<J/\eta<5$ where  the purification time shows quadratic power law dependence
$\eta t_{\mathrm{p},\mathrm{num}}\sim N^2$.
In the following, we derive an approximate evolution equation describing the dynamics of \emph{moments} of density matrices subject to a common measurement protocol. On this basis we will be able to predict the  boundary between the two phases, the symmetries characterizing them, and the mechanisms safeguarding their stability.

\noindent \emph{Diagonal projection ---} The starting point of our construction is
the observation that the random outcome of repeated measurements acts as a  source of
noise suppressing Fock space matrix elements $\rho_{nm}$ off-diagonal in the
measurement basis, $n\not=m$, by decoherence. The stochastic Schr\"odinger equation
formulation of measurement dynamics (see~\cite{supplement}) makes this interpretation
concrete and can be used to derive an effective equation for the  states $P\equiv
\mathcal{P}\rho$, where $\mathcal{P}$ is a projector onto the subspace spanned by the
Fock space diagonal states $\{|n\rangle\langle n|\}$. In the supplemental material we
show that the discrete time dynamics of the diagonal coefficients $P_{n,t}$ is
governed by the evolution equation
\begin{align}
    \label{eq:PEvolution}
      d P_t\equiv P_{t+\delta t}-P_t =  -(\delta_t X_H + V_{\phi_t})P_t,
 \end{align} 
 where  the action of the two operators on the r.h.s is defined through
 \begin{align}
     \label{eq:POperatorDef}
     (X_HP)_n&\equiv \sum_m W_{nm} (P_n-P_m),\cr
     (V_\phi P)_n&\equiv \sum_i 2\phi_i(n_i-\bar n_i)P_n.
 \end{align}
 Here, the first term describes incoherent transitions between different 
 occupation states in Fock space with rates 
 $W_{nm}=2\Gamma^{-1}|H_{nm}|^2$. They are induced by transient fluctuations out of the diagonal state with matrix elements $H_{nm}$, followed by relaxation back into it with a decay rate 
 $\Gamma\sim \eta$. Specifically, for the SYK model  
 $X_H=g\sum_{ijkl} (1-\sigma_{x,i}  \sigma_{x,j}  \sigma_{x,k}\sigma_{x,l})$, with $g=J^2/(\eta N^3)$ where we used $\Gamma=\eta$, and the Pauli matrices $\sigma_{x,i}$ flip the occupation of the occupation of site $i$ according to $0,1\rightarrow 1,0$.

 The second term introduces locally correlated measurement noise, $\langle  \phi_{i,t} \phi_{i',t'}\rangle=\eta\delta t\delta_{t,t'} \delta_{ii'}$. The noise affects states the more the farther they are away from the instantaneous expectation values $\bar  n_i\equiv \langle   n_i\rangle=\sum_n P_n n_{i}$. (The subtraction of the expectation values also safeguards the positivity and
 probability conservation of the diagonal states: $d \sum_n P_n
 =\sum_i 2\phi_i \sum_n P_n (n_i-\overline n_i)=0$.) The self consistent coupling of the r.h.s. of Eq.~\eqref{eq:POperatorDef} to the solution $P$ via the expectation values $\bar n_i$ makes the equation difficult to solve. In the following, we consider cases where these terms are expected to play no significant role. These should include the physics of the weak measurement regime, and at least qualitative aspects of the dynamics across the transition. 

\noindent \emph{Generator of dynamics ---} Our objects of interest are $r$-fold replicated
tensor products $P^{(r)}\equiv \langle P\otimes \ldots \otimes P\rangle$ averaged
over noise ($r=2$ for the averaged purity). Taking the  average is facilitated by the It\^o-discretization of
Eq.~\eqref{eq:PEvolution}, i.e.   $P_t$ depending only on the noise history at
earlier times, $\{\phi_{t'<t}\}$. Passing to a continuum
description, it is then straightforward to derive the master equation~\cite{supplement}
\begin{align}
    \label{HEffective}
    &\partial_t P^{(r)}=-X^{(r)}P^{(r)}\cr
    &\quad X^{(r)}= \sum_a X^a_H-4\eta\sum_{a\not= b}n_i^a n_i^b\equiv X^{(r)}_H+X^{(r)}_M,
\end{align}
where  operators carrying a superscript $a$ act in the $a$th copy of the replica product space.  

The generator $X^{(r)}$ describes a competition between the stochastic hopping dynamics represented by $X_H$, and a tendency to confine the $r$ copies of states to a common configuration of measurement outcomes $\{n_i\}$ (Notice the negative sign in $-4\eta$ which rewards positive correlation in replica space.) Since $X_H$ and $X_M$ appear in the effective Hamiltonian as sums over $\sim N^4$ and $N$ site configurations, respectively, we characterize their relative strength in terms of a parameter $\lambda \equiv (N^4 g)/(N\eta)=N^3 g/\eta$. 

The structural similarity of Eq.~\eqref{HEffective} with an imaginary time Schr\"odinger equation suggests to interpret $P^{(r)}\equiv |P^{(r)}\rangle$ as a state vector with components $P^{(r)}_{n}=\langle n|P^{(r)}\rangle$, $n=(n_1,\ldots,n_r)$, and $X^{(r)}$ as its ``effective Hamiltonian''. At large times,  $|P^{(r)}_t\rangle \to |\Psi^{(r)}_\lambda\rangle$, the physical states will asymptote towards the  measurement strength dependent ground states $|\Psi^{(r)}\rangle$ (the dark states of the replicated Lindbladian measurement dynamics) of $X^{(r)}$. 

To understand the nature of the latter in the regimes  of weak and strong measurement, respectively, it is crucial to note two discrete symmetries of $X^{(r)}$:
The first is 
$\Bbb{Z}_2^{\times r}$ symmetry under $\sigma^{(a)}_{xi}\to (\pm)^{(a)}
\sigma^{(a)}_{xi}$, with a replica-dependent (but $i$-independent) sign factor; this
freedom represents the fermion parity symmetry of each individual of the replicated SYK
systems. The second is a $\Bbb{Z}_2^{\times N}$ symmetry under $\sigma^{(a)}_{zi}\to
(\pm)_i\sigma^{(a)}_{zi}$, with an $i$-dependent (but $a$-independent) sign factor;
this symmetry reflects the physical equivalence of the $2^N$ possible measurement
outcomes. In the following, we discuss how the full symmetry group
$\Bbb{Z}_2^{\times r} \times \Bbb{Z}_2^{\times N}$ is broken by the ground states in the two phases of the system.

\noindent \emph{Replica symmetry breaking transition ---}  In the limiting case of absent measurement, $\lambda\to \infty$, the effective Hamiltonian possesses the $2^r$-fold degenerate ground states,
$|\Psi_\infty^{(r)}\rangle \equiv |\Psi^{(r)}_{s}\rangle =|s_1\rangle\otimes \ldots\otimes |s_r
\rangle$, where $|s\rangle=|\pm\rangle \equiv \otimes_{i=1}^n|\pm\rangle$ with $|\pm
\rangle =\frac{1}{\sqrt{2}}(|0\rangle \pm |1\rangle)$.  These are cat states, fully polarized in  $\pm
x$-direction independently for each replica channel. Their ground state property
follows from the observation that $X_H\equiv g(N^{4}-S^{4}_x)$ affords a
representation in terms of the global spin operator $S_x=\sum_i \sigma_{xi}$. The
$2^r$-fold degeneracy of these states indicates that the weak measurement phase is a
replica symmetry breaking phase. We also note that the ground state is uniformly distributed over Fock space, $|\langle
n^1,\ldots,n^r|\Psi^{(r)}_{s}\rangle|=D^{-r}$, as is typical for quantum ergodic states. Finally, the symmetry breaking is
stable under the  inclusion of weak but finite measurement;  it takes
``thermodynamically many'' $\mathcal{O}(N)$ matrix elements of the measurement
operator to flip one cat state into another.

In the opposite case, $\lambda=0$, we have the $2^N$-fold
degenerate set of ground states $|\Psi^{(r)}_n\rangle=\bigotimes_i
|n_i^{(r)}\rangle$, where $|n_i^{(r)}\rangle=\bigotimes_a |n_i\rangle$ 
are fully  $\pm z$-polarized replica symmetric states,  independently for each site --- a ``real space'' symmetry breaking configuration. However, for arbitrarily weak  $g>0$, only $r$ matrix elements of the operator $X_H$ are required to flip between  states of identical replica polarization but different site configuration. The actual, non-degenerate ground state is an  equal weight superposition $|\Psi^{(r)}_0\rangle \equiv D^{-1}\sum_n |\Psi^{(r)}_n\rangle$ showing  unbroken $\Bbb{Z}_2^{\times r} \times \Bbb{Z}_2^{\times N}$symmetry: 
The combination of measurements and any
residual system dynamics leads to an homogenization of measurement outcomes at large
time scales. In the limit $J\lesssim\eta$, this homogenization can be described perturbatively in $X_H$ by performing a 'Schrieffer-Wolff' transformation of the Lindbladian~\cite{supplement}. It also reveals the perturbative stability of the strong measurement dynamics discussed above for $0<J\lesssim\eta$.

Since the two ground states of the effective theory, $|\Psi^{(r)}_{0,\infty}\rangle$
 have different symmetry, there must be a discrete symmetry breaking phase transition
 at a finite value of $\lambda$. An estimate for the transition threshold is obtained
 by comparison of the expectation values $\langle
 \Psi^{(r)}|X^{(r)}|\Psi^{(r)}\rangle$ in the respective states. We find  that $\langle \Psi^{(r)}_\infty|X^{(r)}|\Psi^{(r)}_\infty\rangle =0$ while
 $\langle
 \Psi^{(r)}_0|X^{(r)}|\Psi^{(r)}_0\rangle= gr N^4 -4\eta  r(r-1) N$, indicating a
 transition in the $r$-replica system at $\lambda=4(r-1)$. 
 With $\lambda=gN^3/\eta= J^2/\eta^2$, and $r=2$, the energy balance suggests a transition  at $J=2\eta$. This prediction   is compatible
 with the numerically observed change in the time dependence
 of  purification in Fig.~\ref{fig:PurLargeJ}. 

 From the  ground states, one may also compute other signatures of the two phases. For example, one may introduce an entanglement cut by partition of $n=(n_A,n_B)$ into two bit-strings of total length $N=N_A+N_B$. Moments of the reduced (diagonal) density matrix are then obtained as $\langle \tr_A(\rho_A^r)\rangle=\sum_{n_A}\sum_{n_{Bi}}\langle (n_A,n_{B_1}),\ldots (n_A,n_{B_r})|\Psi_\lambda\rangle$. A straightforward calculation obtains the entanglement entropies in the two phases, $S_A=\partial_{r}|_{r=1} \langle \tr_A(\rho_A^r)\rangle$ come out as $S_{A,\lambda\gg 1}=N_A\ln 2$ and $S_{A,\lambda\ll 1}=0$. The change from volume law to vanishing entanglement entropy reveals $S_A$ as an alternative indicator of the transition~\cite{Fisher2018,Skinner2019}. However, for the small system sizes considered here, this change is difficult to resolve in simulations.

\noindent\emph{Conclusions --- } The starting point of this paper was the observation
that in the measured quantum dynamics of non-integrable systems the two sources of complexity ``continued measurement'' and ``chaotic
dynamics'' to some degree neutralize each other. 
We exploited this principle to formulate
a comprehensive approach to the description of measurement dynamics for interacting
systems of mesoscopic (number of particles large but finite) extensions. Its elements
included explicit calculations of the purity for strong and weak measurement, and  an analysis of the symmetry breaking transition between them. In view of the growing importance of measured quantum dynamics in mesoscopic (``NISQ'') device structures, various directions of future research present themselves. For example, it would be interesting to extend the theory to systems where local correlations slow the scrambling of information by Lieb-Robinson bounds. It would also be nice to identify a one-does-it-all path integral framework, with account for coherences (required to describe the weak measurement phase), and self consistent update of measurement records (required to describe the strong measurement phase).

\begin{acknowledgments}
We thank M. Gullans and D. Huse for very fruitful discussion.
 T.~M.~acknowledges financial support by Brazilian agencies CNPq
and FAPERJ. We acknowledge support from the Deutsche
Forschungsgemeinschaft (DFG) within the CRC network TR 183 (project grant 277101999)
as part of projects A03 and B02. 
S.D. acknowledges support by the Deutsche Forschungsgemeinschaft (DFG, German Research Foundation) under Germany’s Excellence Strategy Cluster of Excellence Matter and Light for Quantum Computing (ML4Q) EXC 2004/1 390534769, and by the European Research Council (ERC) under the Horizon 2020 research and innovation program, Grant Agreement No. 647434 (DOQS). 
\end{acknowledgments}

\appendix
\section{Derivation of  Eq.~(4) in the main text} 
\label{sub:derivation_of_the_}
\label{sub:derivation_of_the_}

In this section, we derive Eq.(4) in the main text in a succession of two steps. We first investigate the dynamics of individual propagators $P$ for a given realization of the measurement noise, and then consider the average of multiple of these objects.

Our  starting point is the observation that $P_{nmt}$ can be interpreted as the evolution of an initial density operator $ \rho(0)=|m\rangle\langle m|$ to the state $\rho(t)=|n\rangle\langle n|$ by being continuously projected onto configurations diagonal in the measurement basis. To describe the evolution of this object, we consider $\rho$ as a vector in the Fock space tensor product $\mathcal{F}\otimes \mathcal{F}^\ast$ and introduce the projection $P=\mathcal{P}\rho$, where $(X_\mathcal{P})_{nm}=(\mathcal{P}X)_{nm}=\delta_{nm}X_{nn}$ projects on diagonal configurations, and $\mathcal{Q}=\mathrm{id}-\mathcal{P}$ on the complement of off-diagonal ones. 

First consider the Hamiltonian contribution to the time evolution of $\rho$. Defining the ``super-operator'' $X_H \rho\equiv -i [H,\rho]$, we represent the projected von Neumann equation as 
\begin{align}
   d_t \mathcal{P}\rho &= \mathcal{P}X_H \mathcal{Q} \rho,\cr 
   d_t \mathcal{Q}\rho &= \mathcal{Q}X_H \mathcal{P}\rho+ \mathcal{Q}X_H \mathcal{Q} \rho,
\end{align}
where we noted that a  $\mathcal{P}X_H \mathcal{P}$ contribution drops out due to the
commutator structure of $X_H$. We now solve the second equation under one
phenomenological assumption (which can be backed by microscopic calculations for
simple models such as the SYK model): chaotic systems efficiently decohere
off-diagonal density operator matrix elements in generic bases. We thus assume that
$\mathcal{Q}X_H \mathcal{Q} \rho \simeq -\Gamma \mathcal{Q}\rho$, with a
 fast decay rate, $\Gamma$, whose detailed value we leave unspecified.  With this approximation,
$\mathcal{Q}\rho(t) \simeq  t_{\mathrm{p}}^t ds\, e^{-\Gamma (t-s)}\mathcal{Q}X_H
\mathcal{P}\rho(s)\simeq \Gamma^{-1}\mathcal{Q}X_H \mathcal{P} \rho(t)$, and
substitution into the first equation leads to $d_t \mathcal{P}\rho =
\Gamma^{-1}(\mathcal{P}X_H \mathcal{Q})(\mathcal{Q}X_H
\mathcal{P})(\mathcal{P}\rho)$. Translating back to a representation in terms of the
coefficients $P_n$, this becomes
\begin{align}
     d_t P = -X_H P, \qquad (X_H P)_n=\sum_m W_{nm} (P_n-P_m), 
 \end{align} 
 with $W_{nm}=\frac{2}{\Gamma}|H_{nm}|^2$. Notice the structure of a master equation with transition rates $W_{nm}$. This structure preserves the positivity and probability conservation of the distribution $P$.

In contrast to the Hamiltonian, the measurement operator stays within the diagonal
subspace defined by the measurement basis. Its action can be inferred from the
discrete time stochastic Schr\"odinger equation, $d|\psi_t\rangle\equiv |\psi_{t+\delta t}\rangle
-|\psi_{t}\rangle =(-i\delta t H   - \frac{\eta\delta t}{2}\sum_i M_i^2+\sum_i \phi_i M_i) |\psi_{t}\rangle$, where the measurement noise is Gaussian correlated
with $\langle  \phi_{i,t} \phi_{i',t'}\rangle= \delta t\eta
\delta_{t,t'}\delta_{ii'}$, the operator $M_i = O_i - \bar O_i$ contains the
measurement operator self consistently corrected for the instantaneous average $\bar
O_i = \langle \psi |O_i |\psi\rangle=\sum_n O_{i,n} P_n$. A Taylor expansion of the
discrete evolution equation  of $\rho_t =|\psi_t\rangle \langle \psi_t| $ to leading
order in $\delta t$ (including the standard replacement $ \phi_{i,t} 
\phi_{i',t}\to \delta t \eta\delta_{ii'}$) then defines the measurement contribution
to the evolution as
\begin{align*}
      -V_\phi P& = \sum_i\left(\frac{\eta \delta t}{2} (\{M_i^2,P\} - 2M_i P M_i) + \{ \phi_iM_i,P\}\right) \cr 
      &= 2\sum_i \delta \phi_i M_i P.
  \end{align*} 
   Adding the Hamiltonian contribution, we obtain the discrete time evolution equation
\begin{align}
    \label{eq:PEvolution}
      d P_t\equiv P_{t+\delta t}-P_t =  -(\delta tX_H + V_{\phi_t})P_t,
 \end{align} 
 where  $P_{n0}=\delta_{n,m}$. What makes this equation complicated (nonlinear) is the self consistent dependence of $V_\phi$ on the r.h.s. on the solution via the averages $\bar O_i$. However, in the regimes studied in this paper, these values either average out (strong measurement phase), or are intrinsically small (weak measurement phase). We therefore ignore them throughout, and replace $M_i\to O_i$ by the naked measurement operators.  

 We now consider the noise average of the ``tensor product'' $P^{(r)}=\langle P\otimes \ldots \otimes P\rangle$ of $r$ propagators. (It is useful to think about these expressions in a quantum mechanics inspired language: in it, individual $P$'s represent  time dependent ``wave functions'' $P_n(t)=P_{nt}$ with initial values $P_{n0}=\delta_{nm}$, and the above is their $r$-body generalization.) To obtain Eq.~(5) in the main text, we consider the time differences
 $dP^{(r)}_t=\langle \bigotimes P_{t+dt}-\bigotimes P_t\rangle$, and expand to up to
 second order in $d_t$. The first order expansion leads to $-\delta t \sum_{a}X_H^a
 P^{(r)}_t$ the first term in Eq.~(4) in the main text, and the second order one to a
 an operator $4\sum_{a\not=b}\sum_{ij} \langle\phi_{i,t} \phi_{j,t}\rangle O_i^a
 O_j^b\,  P^{(r)}_t$, where we used that  $P_t$ depends only on $\{\phi_{t'<t}\}$ and
 so the average over $\phi_t$ decouples. From the Gaussian correlation of
 $\phi_{i,t}$ we obtain $4\delta t \eta \sum_{a\not=b}\sum_i O^a_i O^b_i P^{(r)}_t$. Adding the Hamiltonian operator dividing by $\delta$ and taking the continuum limit, we obtain Eq.~(4).

\section{Weak measurement} 
\label{sub:weak_measurement}
In this section, we discuss in detail how the average purification is obtained for the protocol of sporadic projective measurements described in the main text. An  initial maximally mixed
configuration, $\rho_0=\frac{1}{D}\mathds{1}$ remains invariant  under
system evolution, until the first measurement of qubit no. $i$ in
state $n_i$. After it, the density operator is given by $\frac{2}{D}P_1$, where $P_1\equiv P_{i}^{n_i}$ projects the $i$th qubit
onto a state $n$, and time evolution under $U_{\Delta t}\equiv U$ up to a time $\Delta t$ defines the state $\rho_1\equiv
\mathcal{N}_1 U_1 P_1U_1^\dagger$, with normalization $\mathcal{N}_1^{-1}=\mathrm{tr}(U_1 P_1 U_1^\dagger)=D/2$. A second projective measurement, now of qubit $j$, collapses this state to $\rho_2 \equiv \mathcal{N}_2
P_2 U_1 P_1U_1^\dagger P_2$, with a second projector $P_2\equiv P_j^{n_j}$, and updated normalization $\mathcal{N}_2^{-1}=\mathrm{tr}(P_2 U_1 P_1U_1^\dagger P_2)$. 

Iteration of this evolution defines the state $\rho_l = \mathcal{N}_l Z_l$ with
$\mathcal{N}_l=\mathrm{tr}(Z_l)$ and $Z_l=P_l U_l P_{l-1} \ldots P_{l-1}U_l^\dagger
P_l$, or $Z_l=P_l U_l Z_{l-1}U_l^\dagger P_l$ in a recursive definition. The purity
after $l$ measurements is given by $\langle
\mathcal{N}_l^2\mathrm{tr}(Z_l^2)\rangle$. We evaluate this expression under the 
simplifying assumptions mentioned in the main text: statistical independence of different $U_k$, and independence of the normalization factor and
the operator trace $\langle \tau\rangle\approx \langle
\mathcal{N}_l^{-2}\rangle^{-1}\langle \mathrm{tr}(Z_l^2)\rangle$. 

The statistics of each update step then reduces to the computation of Haar averages of four matrix elements of $U$ and its adjoint. For a generic set of matrix elements, we have the formula (from now on, $\langle \ldots\rangle$ denotes Haar measure averaging)
\begin{align}
    &\langle U^\pd_{kl }U^\pd_{mn}U^\dagger_{n'm'}U^\dagger_{l'k'}\rangle=\cr 
    &\quad= c_1 \delta_{kk'}\delta_{ll'}\delta_{mm'}\delta_{nn'}+ c_1 \delta_{km'}\delta_{ln'}\delta_{mk'}\delta_{nl'}+\cr 
    &\quad + c_2 \delta_{kk'}\delta_{ln'}\delta_{mm'}\delta_{nl'}+
    c_2 \delta_{km'}\delta_{ll'}\delta_{mk'}\delta_{nn'},
\end{align}
with coefficients $c_1=(D^2-1)^{-1}$ and $c_2=-(D(D^2-1))^{-1}$. 

\noindent\emph{Recursion:} From this general fourth moment it is straightforward to compute the trace averages
\begin{align}
    \label{eq:TrUvsO}
    &\langle \mathrm{tr}(B U A U^\dagger B U A U^\dagger )\rangle=\cr 
    &\quad =\frac{1}{D^2-1}\Big(\mathrm{tr}(B^2)(\mathrm{tr}(A))^2+(\mathrm{tr}(B))^2\mathrm{tr}(A^2)-\cr 
    &\quad -\frac{1}{D}\mathrm{tr}(A^2)\mathrm{tr}(B^2)-\frac{1}{D}(\mathrm{tr}(A)\mathrm{tr}(B))^2\Big)\cr 
    &\langle (\mathrm{tr}(B U A U^\dagger))^2\rangle=\cr 
    &=\frac{1}{D^2-1}\Big((\mathrm{tr}(A))^2 (\mathrm{tr}(B))^2+\mathrm{tr}(A^2)\mathrm{tr}(B^2)-\cr 
    &\quad -\frac{1}{D}(\mathrm{tr}(A))^2\mathrm{tr}(B^2)-\frac{1}{D}(\mathrm{tr}(B))^2\mathrm{tr}(A^2)\Big)
\end{align}
for general matrix operators $A,B$. We now use these expressions to the one measurement step purity update. To this end,
define the two traces $X'_1=\langle(\mathrm{tr}(Z'))^2\rangle$
and $X'_2=\langle\mathrm{tr}(Z^{\prime2})\rangle$ with $Z'\equiv Z_{l-1}$. We now consider the update
relation $Z\equiv Z_l  = P U Z' U^\dagger P$ with $U\equiv U_l$, $P\equiv P_l$ and the updated
traces $X_1=\langle (\mathrm{tr}(Z))^2\rangle=\langle (\mathrm{tr}(P U
Z' U^\dagger P ))^2\rangle$ and $X_2 = \langle  \mathrm{tr}(PU Z' U^\dagger P U Z' U^\dagger)\rangle$. The structure of these traces is covered by the above auxiliary relations. With $\tr(P)=D/2$, a straightforward computation leads to the closed relation
\begin{align}
    X=\left(\begin{matrix}
        z_1&z_2\cr
        z_2&x_1
    \end{matrix} \right)X', 
\end{align}
where $X=(X_1,X_2)^T$, and $z_1=\frac{D}{2}(\frac{D}{2}c_1+c_2)$, $z_2=\frac{D}{2}(\frac{D}{2}c_2+c_1)$. With the matrix eigenvectors $|\pm\rangle$, eigenvalues $(z_1\pm z_2)$, the initial condition $X_0=(1,D^{-1})$ describing a normalized and fully mixed state, and the overlap $\langle \pm|X_0\rangle =\frac{1}{\sqrt{2}}(1+s D^{-1})$, the recursion relation is solved by 
\begin{align}
    X_l=\frac{1}{\sqrt{2}}\sum_{s=\pm}|s\rangle (z_1+s z_2)^l (1+s D^{-1}). 
\end{align}
From here, we obtain the purity of the normalized states as $\langle \tau_l\rangle  = X_{2l}/X_{1l}$ as
\begin{align}
    \langle\tau_l\rangle\approx \frac{(z_1+z_2)^l (1+ D^{-1})-(z_1-z_2)^l (1- D^{-1})}{(z_1+z_2)^l (1+ D^{-1})+(z_1-z_2)^l (1- D^{-1})}
\end{align}
To leading order in an expansion in $D^{-1}$, we have $z_1\pm z_2\approx \frac{1}{4}\pm \frac{1}{D}$. In this approximation, and for large $l\sim t/\eta N\gg 1$ the above expression then simplifies to Eq.~(1) of the main text.
\section{Stability of the two phases} 
\label{sec:stability_of_the_two_phases}

In this section, we discuss the stability of the weak and the strong measurement regimes away from the limits of diverging and vanishing strength parameters, respectively.

\begin{figure}[t!]
\centering
\includegraphics[width=8.5cm]{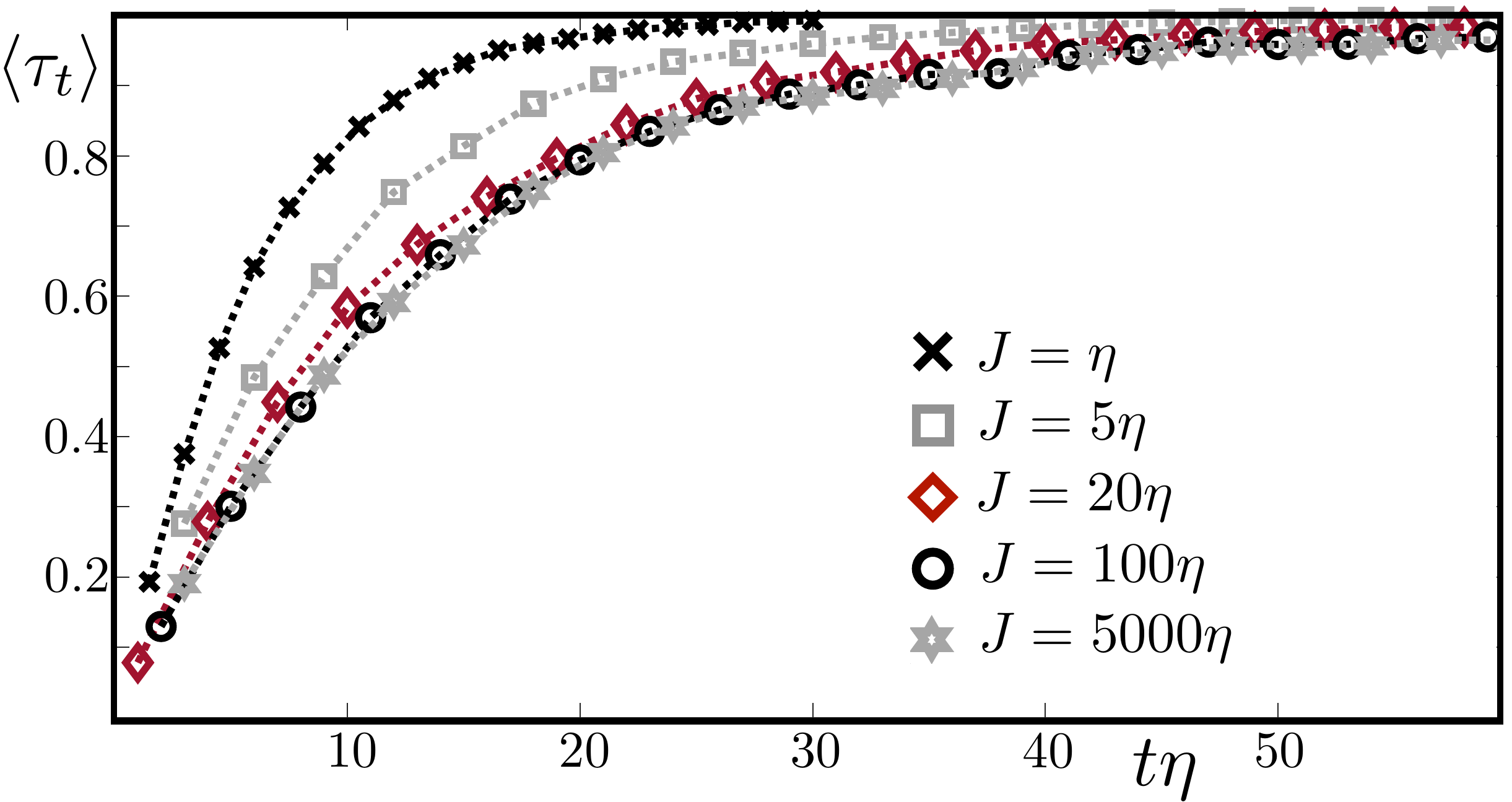}
\caption{\label{fig:LargeJ_Independence}
Purity $\langle\tau_t\rangle=\langle\text{tr}\rho_t^2\rangle$ as a function of time of an initially fully mixed state $\rho_0\sim\mathds{1}$ for $N=8$ qubits and for different values of $J/\eta$. For $J\gg\eta$ the purification dynamics becomes $J$-independent and the curves for larger and larger $J$ converge towards the curve describing the $J=\infty$ limit. In this limit, the evolution is independent of the value of $J$ as predicted by an evolution with independent, Haar-random matrices.}
\end{figure}

\noindent \emph{Weak measurement ---} Evidence for the stability of the weak measurement regime follows from the analysis of the purification time in the main text. The essential condition leading to a $J$-independent time scale $t_\mathrm{p}\sim D/N\eta$ was $J>N\eta$, i.e. full scrambling at time scales shorter than the time between two measurements. The numerical analysis indeed confirms the presence of an extended regime of $J$-independent $t_\mathrm{p}$. This is illustrated in Fig.~\ref{fig:LargeJ_Independence}, where purification trajectories for $J\gg\eta$ (in particular $J\ge20\eta$ for $N=8$ qubits) quickly converge onto a $J$-independent evolution, and corresponding to the limit $J\rightarrow\infty$. In the language of the generator $X^{(r)}$ the same phenomenon is expressed through the stability of the cat ground states $|\Psi^{(r)}_\infty\rangle$. Besides the exponentially long purification time, this state is characterized by a volume law entanglement entropy, $S_a=N_A\ln 2$, where $N_A$ is the bipartition size.

\noindent \emph{Strong measurement ---} In the regime of strong measurement, the dominant generator $X_M^{(r)}$ requires  fast relaxation into any one of the replica homogeneous states $|\Psi_n^{(r)}\rangle$ defined in the main text. Physically, these are states of identical measurement result, $n$, in each replica channel. To understand what happens within the $2^N$-fold degenerate space of these states, we perturbatively include the operator $X_H^{(r)}$ into the picture. More specifically, one may employ a Schrieffer-Wolff transformation to derive an  operator describing virtual transitions out of the measurement ground state induced by $X_H^{(r)}$. For example, for  two replicas, $r=2$, this operator reads $X_\mathrm{SW}^{(2)}= c\frac{g^2}{\eta}\sum_{ijkl} (\sigma_{xi}^1\otimes \sigma_{xi}^2)\ldots (\sigma_{xl}^1\otimes \sigma_{xl}^2)$, with a numerical constant $c$. This operator mixes between different $|\Psi_n^{(2)}\rangle$. Its ground state is the uniform superposition $|\Psi_0^{(2)}\rangle = \frac{1}{D}\sum_n |\Psi_n^{(2)}\rangle$ mentioned in the main text. As long as we are in the regime of perturbative stability of this construction, $J\lesssim \eta$ this replica symmetric configuration should describe the state of the system. Salient features of the construction include fast purification time, $t_\mathrm{d}\simeq \eta^{-1}\ln N$ and vanishing entanglement entropy.

\section{Numerical implementation} 
\label{sec:num_imp}
In order to simulate the purification dynamics numerically, we implement the monitored Lindblad evolution for the complete density matrix $\rho_t$. In each time step, the density matrix evolves under the stochastic master equation $(d\rho_t=\rho_{t+\delta t}-\rho_t)$
\begin{align}
d\rho_t&=-i \delta t [H,\rho_t]-\sum_i \frac{\eta\delta t}{2}[M_i,[M_i,\rho]]+\{\phi_i M_i,\rho_t\}.\nonumber
\end{align}
Here, $H$ is the SYK Hamiltonian and $M_i=n_i-\text{tr}(\rho_t n_i)$ are the measurement operators as provided in the main text. Due to the It\^o calculus, $\phi_i\phi_j=\eta\delta t \delta_{i,j}$, each time step is efficiently implemented by a Trotterized protocol, consisting of a matrix multiplication $\rho_{t+\delta t}= V_{t}U\rho_t U^\dagger V_t$ with $V_t=\exp(-\sum_i \eta\delta t M_i^2-\phi_i M_i)$ and $U=\exp(-iH\delta t)$. The matrix $V_t$ has to be computed in each time step due to the dependence of the operators $M_i$ on $\rho_t$. The computation of $V_t$ and the multiplication with $\rho_t$ can be implemented efficiently since $V_t$ is diagonal in the local Fock state basis. The off-diagonal matrix $U$ is time independent and is computed only once per trajectory from the SYK Hamiltonian $H$. 

In the simulations, we work in units of $\eta=1$, a time discretization of $\delta t=0.05$ and perform averages over $500$ simulated trajectories for each set of parameters. We have tested also smaller time discretizations down to $\delta t=0.001$ for limiting cases of small and large $J$ but no qualitative differences have been observed. For each trajectory, one realization of the SYK Hamiltonian is determined by randomly drawing the couplings $J_{abcd}$ from a Gaussian distribution with zero mean and variance $\langle J_{abcd}^2\rangle=6J^2/(2N)^3$. The trajectory average therefore represents both an average over different measurement outcomes and over SYK realizations. For $N=8$ and $N=10$ we also simulated a limited number of trajectories corresponding to the same SYK Hamiltonian, and we have not observed any significant difference compared to drawing a new Hamiltonian for each trajectory.

\begin{figure}[t!]
\centering
\includegraphics[width=8.5cm]{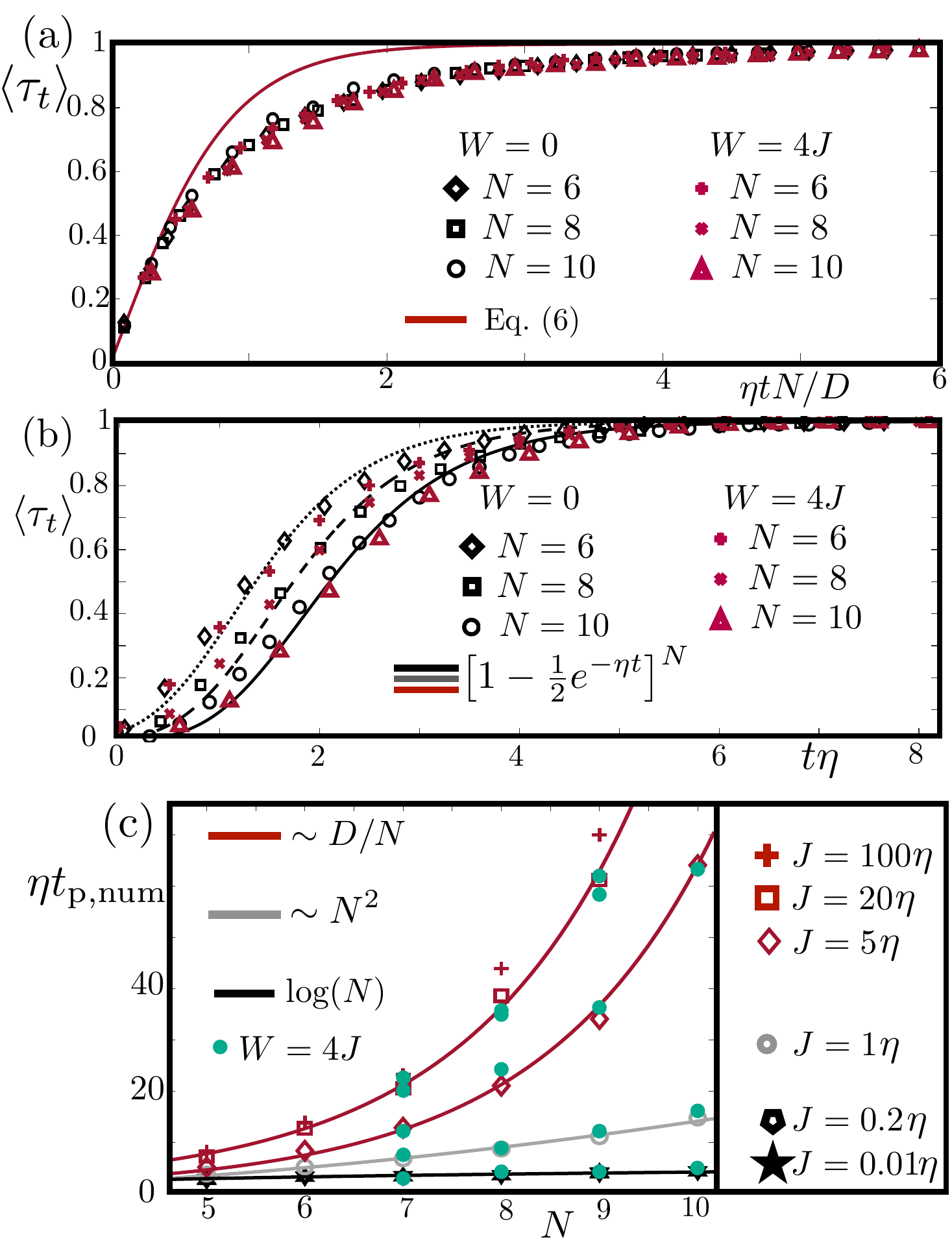}
\caption{\label{fig:Purification_bandwidth}
Purification with additional two-body hopping Hamiltonian $\hat H_2$ with a bandwidth $W=4J$. (a) Time evolution of the purity $\langle\tau_t\rangle=\langle\text{tr}\rho_t^2\rangle$ for an initial maximally mixed state $\rho_0\sim \mathds{1}$ at weak measurement rate ($J=5\times 10^3\eta$) for different system sizes $N$. We compare the purification without two-body Hamiltonian, $W=0$ (black symbols), to the purification with two-body Hamiltonian, $W=4J$ (red symbols). Both cases collapse onto the same trajectory after appropriate rescaling of time by the Hilbert space dimension $D$. The $W=0$ data is identical to Fig.~2 (a) in the main text. (b) Same comparison as in (a) but for strong measurement rate ($J=10^{-2}\eta$). The data for $W=0$ and $W=4$ both match the analytical prediction for $J=0$. (c) Numerically obtained purification times $t_{\text{p,num}}$ for $W=4$ (turquoise) on top of the $W=0$ data from Fig.~2(c) in the main text. No change in the general behavior or on the location of the purification transition is observed (see Fig.~2 in the main text for more details).}
\end{figure}

\section{Purification with additional two-body Hamiltonian}
In order to verify the robustness of the purification dynamics against perturbations,
we consider an additional two-body Hamiltonian, such that $\hat H=\hat H_2+\hat
H_{\text{int}}$, were $\hat H_{\text{int}}$ is the four-body SYK Hamiltonian analyzed
in the main text. For $\hat H_2$ we assume a nearest neighbor hopping Hamiltonian,
which is off-diagonal in the measurement basis and has a bandwidth $W$. For
concreteness, it is given by $\hat H_2=\frac{W}{4}\sum_{l}(c^\dagger_l
c_{l+1}+\text{H.c.})$. In the main text, we examined the behavior for $W=0$, where
the ground state of the system is a non-Fermi liquid. Here, we  look at the case
$W>J$  to verify that  in the opposite Fermi liquid regime, too, the system undergoes a purification transition with the same properties.

We test this limit by numerical simulation of the purification dynamics at a fixed bandwidth $W=4J$. The results are shown in Fig.~\ref{fig:Purification_bandwidth}, where we compare the time evolution for weak measurements Fig.~\ref{fig:Purification_bandwidth}(a), strong measurements (b) and the purification times (c) for the cases $W=0$ and $W=4J$. As seen in the figure, there are no detectable differences in the purification trajectories nor in the purification times in both cases. This leads to the conclusion that the purification is dominated exclusively by the non-integrable part of the Hamiltonian $\hat H_{\text{int}}$.

\bibliography{Biblio}

\end{document}